\documentclass[12pt]{article}

\def\be{\begin{equation}}
\def\ee{\end{equation}}

\begin{document}

\title{Genetic code on the dyadic plane}

\author{A.Yu.Khrennikov, S.V.Kozyrev}

\maketitle

\begin{abstract}
We introduce the simple parametrization for the space of codons
(triples of nucleotides) by $8\times 8$ table. This table (which we
call the dyadic plane) possesses the natural 2--adic ultrametric. We
show that after this parametrization the genetic code will be a
locally constant map of the simple form. The local constancy of this
map will describe degeneracy of the genetic code.

The map of the genetic code defines 2--adic ultrametric on the space
of amino acids. We show that hydrophobic amino acids will be
clustered in two balls with respect to this ultrametric. Therefore
the introduced parametrization of space of codons exhibits the
hidden regularity of the genetic code.
\end{abstract}

\section{Introduction}

Investigation of properties of the genetic code attracts a lot of
interest, see \cite{Swanson}, \cite{Miguel}, \cite{SjoWold}. In the
present paper we apply $p$--adic and ultrametric methods to
investigation of the genetic code. In particular we show that
degeneracy of the genetic code is described by local constancy of
some map defined on some ultrametric space.

Let us remind that an ultrametric space is a metric space where the
metric $d(X,Y)$ satisfies the strong triangle inequality:
$$
d(X,Y)\le {\rm max}\left(d(X,Z),d(Y,Z)\right),\qquad \forall Z
$$
An ultrametric space is a natural mathematical object for
description of a hierarchical system. In particular, any two balls
in ultrametric space either do not intersect or one of the balls
contains the other ball. On ultrametric spaces there exist many
locally constant functions, i.e. functions which are constant on
some vicinity of any point, but not necessarily constant on the
whole space.

The example of ultrametric space is the field of $p$--adic numbers,
which is the completion of the field of rational numbers with
respect to $p$--adic norm $|x|_p$, defined as follows: for rational
number $x=p^{\gamma}{m\over n}$ its norm $p$--adic is
$$
|x|_p=p^{-\gamma}
$$

Starting from purely mathematical ideas and applications to high
energy physics and string theory, methods of $p$--adic mathematical
physics \cite{VVZ}, \cite{Andr} were developed. Ultrametric methods
found applications in physics of complex systems. The first
application of ultrametricity to physics was the description replica
symmetry breaking in theory of spin glasses \cite{MPV}. In the
following $p$--adic parametrization for the Parisi matrix used in
the theory of spin glasses was constructed \cite{ABK}, \cite{PaSu}.

Applications of $p$--adic numbers to information theory and, in
particular, to description of cognitive processes and complex social
systems were described in books \cite{Andr}, \cite{Andr3}. These
application are based on possibility to describe hierarchical
structure of information (for example, in psychology) with the help
of $p$--adic metric or more complex ultrametric.

In papers \cite{Andr4}, \cite{Andr5} it was proposed to use the
4--adic information space (in the sense of \cite{Andr},
\cite{Andr3}) for representation of genetic information. In
particular, the problem of degeneracy of genetic code was
considered. The 4--adic polynomial dynamical systems were used. In
this approach every amino acid correspond to some cycle of the
dynamical system in the space of codons (in the 4--adic
representation), the detailed exposition can be found in
\cite{Andr6}.

Let us remind that the genetic code is the map from the space of
codons (triples of nucleotides) onto the space of amino acids (plus
the additional stop--codon Ter). There is a well known problem of
degeneracy of the genetic code: there are 64 codons, but the number
of amino acids (including the stop--codon) is 21.

A model to describe the degeneracy of genetic code using the quantum
algebra $\mathcal{U}_q (sl(2)\oplus sl(2))$ in the limit $q \to 0$,
was proposed in \cite{sorba}, \cite{sorba1}. In these papers the
analogy between the genetic code and quark models of barions was
discussed.

In paper  \cite{DD} by B.Dragovich and A.Dragovich the information
5--adic space was used to describe the degeneracy of genetic code by
clustering in the corresponding ultrametric. The new approach was
proposed --- the genetic code was represented by the locally
constant map defined on the ultrametric space of codons and
degeneracy of the genetic code was described by local constancy of
the map.

In the present paper we parameterize the space of codons by the
$8\times 8$ table (the dyadic plane), and represent the degeneracy
of the genetic code by local constancy of the map, defined on the
dyadic plane. Our construction differs from the construction of
\cite{DD} by different parametrization of the space of codons: we
use for description of the space of codons the 2--dimensional
object. Also, our construction of ultrametric is different.

Moreover, we show, that the introduced here construction of
ultrametric is related to physical and chemical properties of amino
acids. The genetic code maps the ultrametric on the dyadic plane
(the space of codons) onto the space of amino acids. We show that,
with respect to this ultrametric, hydrophobic amino acids will be
clustered in two balls. Therefore physical properties of amino acids
are related to the considered in the present paper parametrization
of the genetic code.

Let us consider different variations of genetic code. We will see
that the corresponding maps of the dyadic plane onto the space of
amino acids possess different degree of regularity (i.e. the
different character of local constancy). We can conjecture that more
regular maps correspond to older forms of genetic code (since it is
more probable that evolution goes with symmetry breaking).

Thus the introduced in the present paper parametrization of the
space of codons by the dyadic plane exhibits the regular structure
of the genetic code.

Ultrametric was widely used in bioinformatics in construction of
phylogenetic trees starting from nucleotide sequences, see
\cite{bioinformatics}. The results of the present paper show that
one can apply ultrametric methods to investigation of the genetic
code starting from the level of single codons. In particular, one
can use this observation to modify the metric used in computational
genomics.

\section{The genetic code}

In the present Section we put discussion of the genetic code. This
material can be found in \cite{Watson}. The genetic code is the map
which gives the correspondence between codons in DNA and amino
acids. Codon is a triple of nucleotides, nucleotides are of four
kinds, denoted by C, A, T, G (Cytosine, Adenine, Thymine, Guanine),
in total we have 64 codons. In RNA Thymine is replaced by Uracil,
denoted by U.

We have 20 amino acids:  alanine, threonine, glycine, proline,
serine, aspartic acid, asparagine, glutamic acid, glutamine, lysine,
histidine, arginine, tryptophan, tyrosine, phenylalanine, leucine,
methionine, isoleucine, valine, cysteine, denoted correspondingly by
Ala, Thr, Gly, Pro, Ser, Asp, Asn, Glu, Gln, Lys, His, Arg, Trp,
Tyr, Phe, Leu, Met, Ile, Val, Cys,  and the stop--codon Ter.

Genetic code put into correspondence to a codon $C_1C_2C_3$ (where
$C_i=C,A,U,G$) an amino acid or Ter (a stop codon).

There exist several variations of genetic code. Different variants
of genetic code generally coincide but can differ on few codons. The
following two tables describe the vertebral mitochondrial code and
the standard, or eucaryotic, code.

\bigskip

\noindent
\begin{tabular}{|c|c|c|c|}
 \hline \  & \  &  \ & \ \\
    AAA Lys  &  UAA Ter &  GAA Glu &  CAA Gln  \\
    AAU Asn  &  UAU Tyr &  GAU Asp &  CAU His  \\
    AAG Lys  &  UAG Ter &  GAG Glu &  CAG Gln  \\
    AAC Asn  &  UAC Tyr &  GAC Asp &  CAC His  \\
 \hline \  & \  & \  &   \\
    AUA Met  &  UUA Leu &  GUA Val &  CUA Leu \\
    AUU Ile  &  UUU Phe &  GUU Val &  CUU Leu \\
    AUG Met  &  UUG Leu &  GUG Val &  CUG Leu \\
    AUC Ile  &  UUC Phe &  GUC Val &  CUC Leu \\
 \hline \ & \   & \  &   \\
    AGA Ter  &  UGA Trp &  GGA Gly &  CGA Arg  \\
    AGU Ser  &  UGU Cys &  GGU Gly &  CGU Arg  \\
    AGG Ter  &  UGG Trp &  GGG Gly &  CGG Arg  \\
    AGC Ser  &  UGC Cys &  GGC Gly &  CGC Arg  \\
 \hline \ & \ & \ & \\
    ACA Thr  &  UCA Ser &  GCA Ala &  CCA Pro  \\
    ACU Thr  &  UCU Ser &  GCU Ala &  CCU Pro  \\
    ACG Thr  &  UCG Ser &  GCG Ala &  CCG Pro  \\
    ACC Thr  &  UCC Ser &  GCC Ala &  CCC Pro  \\
\hline
\end{tabular}

\bigskip
{Table 1 : The vertebral mitochondrial code}
\bigskip

\bigskip

\noindent
\begin{tabular}{|c|c|c|c|}
 \hline \  & \  &  \ & \ \\
  AAA Lys  &  UAA Ter &  GAA Glu &  CAA Gln  \\
  AAU Asn  &  UAU Tyr &  GAU Asp &  CAU His  \\
  AAG Lys  &  UAG Ter &  GAG Glu &  CAG Gln  \\
  AAC Asn  &  UAC Tyr &  GAC Asp &  CAC His  \\
 \hline \  & \  & \  &   \\
  AUA Ile  &  UUA Leu &  GUA Val &  CUA Leu \\
  AUU Ile  &  UUU Phe &  GUU Val &  CUU Leu \\
  AUG Met  &  UUG Leu &  GUG Val &  CUG Leu \\
  AUC Ile  &  UUC Phe &  GUC Val &  CUC Leu \\
 \hline \ & \   & \  &   \\
  AGA Arg  &  UGA Ter &  GGA Gly &  CGA Arg  \\
  AGU Ser  &  UGU Cys &  GGU Gly &  CGU Arg  \\
  AGG Arg  &  UGG Trp &  GGG Gly &  CGG Arg  \\
  AGC Ser  &  UGC Cys &  GGC Gly &  CGC Arg  \\
 \hline \ & \ & \ & \\
  ACA Thr  &  UCA Ser &  GCA Ala &  CCA Pro  \\
  ACU Thr  &  UCU Ser &  GCU Ala &  CCU Pro  \\
  ACG Thr  &  UCG Ser &  GCG Ala &  CCG Pro  \\
  ACC Thr  &  UCC Ser &  GCC Ala &  CCC Pro  \\
\hline
\end{tabular}

\bigskip
{Table 2 : The eucaryotic code}
\bigskip

\section{Parametrization of the set of codons by the dyadic plane}

In the present Section we introduce the parametrization of the space
of codons by the dyadic plane.

The first step of this construction is the parametrization of the
set of nucleotides by pairs of digits $(x,y)$: 00, 01, 10, 11. This
parametrization is described by the following $2\times 2$ table:
\be\label{22table}
\begin{array}{|c|c|}\hline
A & G \cr\hline U & C \cr\hline
\end{array}=\begin{array}{|c|c|}\hline 00 & 01
\cr\hline 10 & 11 \cr\hline
\end{array}
\ee This parametrization of the set of nucleotides was used in
\cite{Swanson}, \cite{Miguel}, where the Gray code model for the
genetic code was considered. It was mentioned that, since the
nucleotides ${\rm A}=(0,0)$ and ${\rm G}=(0,1)$ are purines, ${\rm
U}=(1,0)$ and ${\rm C}=(1,1)$  are pyrimidines, the different first
digits in the binary representation corresponds to the different
chemical types of the nucleotides. Namely, the nucleotide $(x,y)$
with $x=0$ is a purine, and the nucleotide $(x,y)$ with $x=1$ is a
pyrimidine.

The second digit $y=0,1$ in the considered parametrization
\cite{Miguel} also has the physical meaning. It describes the
$H$--bonding character (weak for $y=0$ and strong for $y=1$).

The second step of our construction is to find parametrization of
the space of codons, using the above parametrization of the set of
nucleotides. To do this we take into account the importance of the
nucleotides in the codon, described by the following rule
\cite{Swanson} \be\label{213} 2>1>3 \ee This means that the most
important nucleotide in the codon is the second, and the less
important nucleotide is the third.

The main idea of the present paper is to combine the parametrization
of nucleotides by $2\times 2$ table and the above order of
nucleotides in the codon and obtain the parametrization of the space
of codons by $8\times 8$ table (the dyadic plane).

We call the dyadic plane the square $8\times 8$, which has the
structure of the group $Z/8Z\times Z/8Z$ (i.e. of the direct sum of
two groups of residues modulo 8). Elements of this group we denote
$(x,y)$:
$$
x=(x_0x_1x_2)=x_0+2x_1+4x_2,\quad y=(y_0y_1y_2)=y_0+2y_1+4y_2,\quad
x_i,y_i=0,1
$$
One can say that $x$ and $y$ in this formula are integer numbers
from 0 to 8 in the binary representation.

Let us construct the correspondence $\rho$ between the dyadic plane
and the set of codons. Using the rule (\ref{213}), we put into
correspondence to the most important (the second) nucleotide in the
codon the largest scale of the $8\times 8$ dyadic plane
--- the pair $(x_0,y_0)$, we correspond to the first nucleotide in the codon the
pair $(x_1,y_1)$, and the third nucleotide in the codon will
determine the pair $(x_2,y_2)$. The nucleotides define the
corresponding pairs $(x_i,y_i)$ according to the rule
(\ref{22table}). We get for the codon $C_1C_2C_3$ the following
representation by the pair of triples of 0 and 1, which we consider
as an element of the dyadic plane:
$$
\rho: C_1C_2C_3\mapsto (x,y)=(x_0x_1x_2,y_0y_1y_2)
$$

Then we enumerate the lines and the columns of the dyadic plane as
follows (in analogy to the $p$--adic parametrization of the Parisi
matrix \cite{ABK}):
$$
\eta: x\mapsto\widetilde{x},\qquad y\mapsto\widetilde{y};
$$
$$
\eta: x_0+2x_1+4x_2 \mapsto 1+ 4x_0+2x_1+x_2;
$$
$$
\eta: y_0+2y_1+4y_2 \mapsto 1+ 4y_0+2y_1+y_2.
$$

Equivalently, we consider the one to one correspondence of numbers
of the lines or columns in the dyadic plane:
$$
\eta: 0, 4, 2, 6,     1, 5,   3, 7 \mapsto 1,2,3,4,5,6,7,8.
$$

After the map $\eta$ the table $8\times 8$ of codons on the dyadic
plane will take the form:

 {
$$
\begin{array}{|c|c|c|c|c|c|c|c|}

\hline AAA & AAG & GAA & GAG & AGA & AGG & GGA & GGG \cr

\hline AAU & AAC & GAU & GAC & AGU & AGC & GGU & GGC \cr

\hline UAA & UAG & CAA & CAG & UGA & UGG & CGA & CGG \cr

\hline UAU & UAC & CAU & CAC & UGU & UGC & CGU & CGC \cr

\hline AUA & AUG & GUA & GUG & ACA & ACG & GCA & GCG \cr

\hline AUU & AUC & GUU & GUC & ACU & ACC & GCU & GCC \cr

\hline UUA & UUG & CUA & CUG & UCA & UCG & CCA & CCG \cr

\hline UUU & UUC & CUU & CUC & UCU & UCC & CCU & CCC \cr

\hline\end{array}
$$
}

The dyadic plane (and, correspondingly, the space of codons)
possesses the 2--dimensional 2--adic ultrametric, which reflect the
rules (\ref{22table}), (\ref{213}): \be\label{ultrametric}
d(C_1C_2C_3,C'_1C'_2C'_3)={\rm max}(|x-x'|_2,|y-y'|_2) \ee
$$
(x,y)=\rho(C_1C_2C_3),\quad (x',y')=\rho(C'_1C'_2C'_3)
$$
This 2--adic norm can take values 1, 1/2, 1/4.

The difference of the introduced here parametrization of the space
of codons by the dyadic plane and the construction of \cite{DD} is
that in \cite{DD} the space of codons was parametrized by
one--dimensional parameter with 5--adic norm, and the rule
(\ref{213}) was not taken into account (the formal ordering of the
nucleotides in the codon $1>2>3$ was used instead).

In papers \cite{Swanson}, \cite{Miguel} the rules (\ref{22table}),
(\ref{213}) were used for the investigation of the genetic code but
the combination of these rules was different from the considered in
the present paper --- instead of ultrametric parametrization the
Gray code model was used.

\section{Genetic code on the dyadic plane}

In the present Section we discuss the vertebral mitochondrial code,
which looks more regular in our approach.

The genetic code in the considered parametrization put into
correspondence to elements of the dyadic plane the amino acids (and
the stop--codon Ter). In this way we obtain for the Vertebrate
Mitochondrial Code the following table of amino acids on the dyadic
plane:

{\Large
$$
\begin{array}{|c|c|c|c|}
\hline \begin{array}{c} {\rm Lys}  \cr \hline{\rm Asn}
\end{array}  & \begin{array}{c}
{\rm Glu}  \cr \hline{\rm Asp}
\end{array}  & \begin{array}{c}
{\rm Ter}  \cr \hline{\rm Ser}
\end{array}  & {\rm Gly}\cr

\hline\begin{array}{c} {\rm Ter}  \cr \hline{\rm Tyr}
\end{array}  & \begin{array}{c}
{\rm Gln}  \cr \hline{\rm His}
\end{array}  & \begin{array}{c} {\rm Trp}  \cr
\hline{\rm Cys}
\end{array}  & {\rm Arg}\cr

\hline \begin{array}{c} {\rm Met}  \cr \hline{\rm Ile}
\end{array}
 & {\rm Val} & {\rm Thr} & {\rm Ala}    \cr
\hline
\begin{array}{c}
{\rm Leu}  \cr \hline{\rm Phe}
\end{array}

 & {\rm Leu}  & {\rm Ser}
 & {\rm Pro} \cr \hline
\end{array}
$$
}

Each small square of this table corresponds is the image (with
respect to the genetic code) of a square $2\times 2$ from the table
of codons. For example, we have the following correspondence
$$
\begin{array}{|c|c|}\hline
AAA & AAG \cr\hline AAU & AAC \cr\hline
\end{array}\to\begin{array}{|c|}\hline {\rm Lys}  \cr \hline{\rm Asn}\cr \hline
\end{array}\, ,\qquad
\begin{array}{|c|c|}\hline
CCA & CCG \cr\hline CCU & CCC \cr\hline
\end{array}\to\begin{array}{|c|}\hline {\rm Pro}\cr \hline
\end{array}
$$

Some of the $2\times 2$ squares in the table of codons map onto one
amino acid (which gives degeneracy 4 for the genetic code). Some of
the squares map onto two amino acids: the first line of the $2\times
2$ square maps onto one amino acid, the second line maps onto the
other amino acid, giving degeneracy 2 for the genetic code. We also
have three cases of additional degeneracy. For example, the second
square in the last line of the table above as well as the upper half
of the first square in the last line map onto the amino acid Leucine
(Leu).

2--Adic balls with respect to the considered above 2--adic norm on
the plane look as follows. All the table is the ball of diameter 1.
A quadrant (quarter of the table), such as for example the right
lower quadrant containing the amino acids Pro, Ser, Thr, Ala is a
ball of diameter 1/2. A square of 4 codons (quarter of quadrant),
say the square containing the amino acid Pro is a ball of diameter
1/4. Finally, any codon can be considered as a ball of zero
diameter.

The major part of degeneracy of the genetic code (besides the
mentioned three cases of additional degeneracy) has the clear
2--adic meaning on the dyadic plane. First, the genetic code map is
always locally constant on the horizontal coordinate $y$ with the
diameter of local constancy $1/4$, and is locally constant on the
half of space of codons on the vertical coordinate  $x$ with the
diameter of local constancy $1/4$. Second, sets with different
character of local constancy are distributed on the dyadic plane in
the symmetric way: the lower right quadrant corresponds to local
constancy with diameter $1/4$ both on $x$ and $y$, the higher left
quadrant corresponds to local constancy with diameter $1/4$ on $y$
(but not on $x$), and the other two quadrants have similar
distribution of squares with different character of local constancy.

We will say that the degeneracy of the genetic code satisfies the
{\it principle of proximity} --- similar codons are separated by
small 2--adic distances on the dyadic plane. Here similarity means
that the corresponding codons encode the same amino acid. We will
see that the principle of proximity has more general application and
is also related to physical--chemical properties of the amino acids.

Let us discuss our choice of the parametrization for the genetic
code. Considering the vertebral mitochondrial code, it is easy to
see that we always have degeneracy of the genetic code on the third
nucleotide. Moreover, this degeneracy always have the same form
--- it is always possible to change in the third nucleotide C by U, and to change A by G.
Also, on the half of the space of codons
we will have complete degeneracy of the genetic code on the last
nucleotide.

Using the proximity principle we describe this degeneracy by local
constancy of the map of the genetic code on small distances.
Moreover, we will describe different (double and quadruple)
degeneracy as a degeneracy of the map with the domain in two
dimensional ultrametric space over one or two coordinates.

In this way we arrive to the map similar to the described above map
$\rho$ of the space of codons onto the dyadic plane, where the third
nucleotide in the codon corresponds to the smallest scale on the
dyadic $8\times 8$ plane (where we have three scales of distance
--- 1, 1/2 and 1/4).

We have to put the other two scales on dyadic plane into
correspondence to the other two nucleotides in codon. We see that if
we correspond to the first nucleotide in the codon the second
(intermediate) scale on the dyadic plane, then the lower right
quadrant will contain four squares with degeneracy four, and the
upper left quadrant will contain eight half--squares with degeneracy
two. Therefore the table of amino acids on the dyadic plane will
have highly symmetric form. We have fixed the form of the map $\rho$
using the local constancy and symmetry for the genetic code.

One could suggest to use for description of the genetic code the
three dimensional dyadic space. Using the described above picture of
degeneracy of the genetic code, we see that for the three
dimensional parametrization of the space of codons it would be
natural to expect 2--times, 4--times and 8--times degeneracy of the
genetic code, which will correspond to the local constancy of the
map on small distances on one, two and three coordinates. But
8--times degeneracy of the genetic code does not exist. Thus the
most natural object for parametrization of the genetic code should
be two dimensional, and we arrive to the dyadic $8\times 8$ plane.

\bigskip

\noindent{\bf Remark}\qquad The eucaryotic code differs from the
vertebrate mitochondrial code by changing of the code for the codons
AGA and AGG, for AUA, and for UGA. Compared to the eucaryotic code,
the vertebrate mitochondrial code corresponds to the simpler and
more regular table, since the corresponding map of codons onto amino
acids possess larger areas of local constancy with respect to the
distance on the dyadic plane. One can conjecture that the vertebrate
mitochondrial code is more ancient than the eucaryotic code, since
evolution with higher probability goes in the direction of breaking
of symmetry.

\section{Physical--chemical regularity of the genetic code}

The map of the genetic code transfers the ultrametric on the dyadic
plane onto the space of amino acids. We define the distance $D(A,B)$
between two amino acids $A$ and $B$ as the minimum of distances
between their preimages in the dyadic plane:
$$
D(A,B)={\rm min}\, d\left(G^{-1}(A),G^{-1}(B)\right)
$$
where $G$ is the genetic code, i.e. the map of the dyadic plane onto
amino acids, $d$ is the ultrametric (\ref{ultrametric}) on the
dyadic plane. For example, distance between His and Gln is equal to
1/4, distance between Pro and Ala is equal to 1/2, distance between
Asp and Ser is equal to 1. These examples show that ultrametric
differs considerably from the Euclidean distance. For example, Asp
and Ser, situated in the neighbor squares, have the maximal distance
between them. This can be discussed as follows: ultrametric distance
between the points is related to the hierarchy of balls containing
these points. Codons corresponding to Asp and Ser lie in the balls
which are far in the hierarchy.

A natural question arise --- does this ultrametric make any physical
sense? We claim that this ultrametric is related to physical
properties of amino acids. Let us discuss the property of
hydrophobicity. This property is related to polarity of the molecule
and its charge in the solvent: hydrophobic molecules are neutral and
non--polar. Hydrophobic amino acids, which are Leu, Phe, Ile, Met,
Val, Cys, Trp, have high probability to be situated inside the
protein (in the hydrophobic kernel), while the hydrophilic amino
acids have high probability to lie on the surface of the protein and
have a contact with water, see the book \cite{FP}.

In the table below we put only hydrophobic amino acids and omit all
the other. It is easy to see that hydrophobic amino acids are
concentrated in the two balls --- the lower left quadrant (Leu, Phe,
Ile, Met, Val) and the third square of the second line (Cys, Trp).

{\Large
$$
\begin{array}{|c|c|c|c|}
\hline \begin{array}{c} {~~~}  \cr \hline{~~~}
\end{array}  & \begin{array}{c}
{~~~}  \cr \hline{~~~}
\end{array}  & \begin{array}{c}
{~~~}  \cr \hline{~~~}
\end{array}  & {~~~~}\cr

\hline\begin{array}{c} {~~~}  \cr \hline{~~~}
\end{array}  & \begin{array}{c}
{~~~}  \cr \hline{~~~}
\end{array}  & \begin{array}{c} {\rm Trp}  \cr
\hline{\rm Cys}
\end{array}  & {~~~}\cr

\hline \begin{array}{c} {\rm Met}  \cr \hline{\rm Ile}
\end{array}
 & {\rm Val} & {~~~} & {~~~}    \cr
\hline
\begin{array}{c}
{\rm Leu}  \cr \hline{\rm Phe}
\end{array}

 & {\rm Leu}  & {~~~}
 & {~~~~~} \cr \hline
\end{array}
$$
}

We see that the property of hydrophobicity is related to 2--adic
norm on the dyadic plane. We say that the introduced parametrization
satisfies the {\it proximity principle} --- ultrametrically close
amino acids have similar physical--chemical properties. Using the
terminology of \cite{Andr}, \cite{Andr3}, one can say that proximity
in ultrametric information space induce similarity of chemical
properties (and, moreover, for hydrophobic amino acids, arrangement
inside the protein, i.e. proximity in the physical space).

The next table contains polar amino acids. We see that their
arrangement satisfies the proximity principle, in particular, all
the seven amino acids in the upper left quadrant are polar.

{\Large
$$
\begin{array}{|c|c|c|c|}
\hline \begin{array}{c} {\rm Lys}  \cr \hline{\rm Asn}
\end{array}  & \begin{array}{c}
{\rm Glu}  \cr \hline{\rm Asp}
\end{array}  & \begin{array}{c}
{\rm Ter}  \cr \hline{\rm Ser}
\end{array}  & {~~~}\cr

\hline\begin{array}{c} {\rm Ter}  \cr \hline{\rm Tyr}
\end{array}  & \begin{array}{c}
{\rm Gln}  \cr \hline{\rm His}
\end{array}  & \begin{array}{c} {~~~}  \cr
\hline{~~~}
\end{array}  & {\rm Arg}\cr

\hline \begin{array}{c} {~~~}  \cr \hline{~~~}
\end{array}
 & {~~~} & {\rm Thr} & {~~~}    \cr
\hline
\begin{array}{c}
{~~~}  \cr \hline{~~~}
\end{array}

 & {~~~}  & {\rm Ser}
 & {~~~} \cr \hline
\end{array}
$$
}

\bigskip\bigskip

\noindent{\bf Acknowledgments}\qquad The authors would like to thank
I.V.Volovich, B.Dragovich and A.Dragovich for fruitful discussions
and valuable comments. This paper has been partially supported by
the grant of The Swedish Royal Academy of Sciences on collaboration
with scientists of former Soviet Union, by the grant of The
V\"axj\"o University on mathematical modeling in physical and
cognitive sciences, by the grant DFG Project 436 RUS 113/809/0-1, by
the grant of The Russian Foundation for Basic Research (project RFFI
05-01-04002-NNIO-a). One of the authors (S.K.) was partially
supported by The Russian Foundation for Basic Research (project
05-01-00884-a), by the grant of the President of Russian Federation
for the support of scientific schools NSh 6705.2006.1 and by the
Program of the Department of Mathematics of Russian Academy of
Science ''Modern problems of theoretical mathematics''.

\end{document}